\newif\ifarxiv
\arxivtrue
\ifarxiv
\pdfoutput=1
\fi

\documentclass{sig-alternate}

\usepackage[T1]{fontenc}
\usepackage[utf8]{inputenc}

\usepackage[british]{babel}
\usepackage[babel]{csquotes}
\MakeAutoQuote{“}{”}
\MakeAutoQuote*{‘}{’}

\usepackage{makeidx}  
\usepackage{times,amsmath,epsfig}
\usepackage{amsmath}
\ifarxiv
\usepackage{lstsemantic}  
\else
\usepackage{../../sty/clange/lstsemantic}  
\fi
\usepackage{color}
\usepackage{rotating}
\usepackage{url}

\usepackage{subfigure}

\usepackage{amssymb}
\usepackage{graphicx}
\ifarxiv\else
\graphicspath{{images/}{../../images/}}
\fi
\usepackage{float}

\usepackage[backend=bibtex,hyperref=auto,
firstinits=true,style=numeric,
urldate=iso8601,isbn=false,doi=false]{biblatex}
\ifarxiv\else
\fi
\DeclareNameAlias{author}{last-first}
\DeclareNameAlias{editor}{last-first}
\DeclareNameAlias{translator}{last-first}
\DeclareFieldFormat{labelnumberwidth}{#1.}
\DeclareFieldFormat{title}{#1}
\DeclareFieldFormat
  [article,inbook,incollection,inproceedings,patent,thesis,unpublished]
  {title}{#1}
\DeclareFieldFormat{journaltitle}{#1}
\newbibmacro*{volume+number+eid}{%
  \printfield{volume}%
  \iffieldundef{number}
    {}
    {(%
      \printfield{number}%
     )%
    }%
  \setunit{\addcomma\space}%
  \printfield{eid}}
\DeclareFieldFormat{url}{\url{#1}}

 \usepackage[hide]{ed}

\usepackage{enumitem}

\begin{document}

\ifarxiv
\copyrightetc{Copyright is held by the author/owner(s).\\
SEMANTiCS, September 4–5 2014, Leipzig, Germany}
\else
\conferenceinfo{SEMANTiCS '14}{September 04--05 2014, Leipzig, Germany}
\CopyrightYear{2014} 
\crdata{978-1-4503-1972-0/13/09}  
\fi

\title{Representing Dataset Quality Metadata\newline using Multi-Dimensional Views}

\numberofauthors{3} 
%
\author{
\alignauthor
Jeremy Debattista\\
       \affaddr{University of Bonn / Fraunhofer IAIS, Germany}\\
       \email{jeremy.debattista@iais-extern.fraunhofer.de} 
\alignauthor
Christoph Lange\\
       \affaddr{University of Bonn / Fraunhofer IAIS, Germany}\\
       \email{math.semantic.web\allowbreak @gmail.com}
\alignauthor
S\"{o}ren Auer\\
       \affaddr{University of Bonn / Fraunhofer IAIS, Germany}\\
       \email{auer@cs.uni-bonn.de}
}

\maketitle
\begin{abstract}
Data quality is commonly defined as \emph{fitness for use}. 
The problem of identifying quality of data is faced by many data consumers.
Data publishers often do not have the means to identify quality problems in their data.
To make the task for both stakeholders easier, we have developed the Dataset Quality Ontology (daQ).
daQ is a core vocabulary for representing the results of quality benchmarking of a linked dataset.
It represents quality metadata as multi-dimensional and statistical observations using the Data Cube vocabulary.
Quality metadata are organised as a self-contained graph, which can, e.g., be embedded into linked open datasets.
We discuss the design considerations, give examples for extending daQ by custom quality metrics, and present use cases such as analysing data versions, browsing datasets by quality, and link identification.
We finally discuss how data cube visualisation tools enable data publishers and consumers to analyse better the quality of their data.
\end{abstract}





\section{Introduction}
\label{sec:introduction}
There are various definitions for the term “Data Quality”.
Robert Pirsig defines quality as \emph{the result of care}~\cite{Pirsig1974-Zen}, whilst Juran defines quality as \emph{fitness for use}~\cite{Juran1974:biblatex}.
Juran's views on data quality were shared by Phillip Crosby, who defined quality as \emph{conformance to requirements}~\cite{Crosby1979}. 
A substantial amount of Linked (Open) Datasets have already been published\footnote{Some statistics can be found in \url{http://lod-cloud.net} and \url{http://stats.lod2.eu}}.
Most of these facts are extracted from heterogeneous sources, including semi-structured data and unstructured data, which causes great variance in quality.
Therefore such sources could lead to various problems such as inconsistencies and incompleteness, which could render a dataset to not be fit for a certain tasks.
Moreover datasets might also evolve during their life-span, leading to increase or decrease in quality.

Identifying the right quality factors for a dataset is a challenge which is always faced by many data consumers.
The main problem stems from the fact that different domains require different quality metrics.
Various research work~\cite{Bizer2008:PhDThesis:biblatex,Flemming2011,Hogan2012:LDC} defines a number of quality factors pertinent to linked open datasets.
Zaveri et al.~\cite{Zaveri2012:LODQ} provide a systematic review of this and further literature, categorising the different metrics.

To enable the definition and use of different metrics in a standardised manner, we introduce the \emph{Dataset Quality Ontology (daQ)}.
daQ is a light-weight core vocabulary for representing the results of quality benchmarking of a linked dataset, again as linked data.
In linked \emph{open} data settings this allows for embedding quality metadata into datasets, thus “stamping” them with a number of quality measures, allowing for the expression of concrete, tangible values that represent the quality of the data.

This paper is an extended version of the original presentation of an early version of daQ~\cite{DebattistaEtAl:daQ:LDOW:2014}.
The main \emph{new} contribution lies in reusing the W3C Data Cube Vocabulary~\cite{w3c:REC-vocab-data-cube-20140116}, which allows for representing data quality metrics as statistical observations in multi-dimensional spaces.
This approach enables the representation and analysis of quality assessments in a fine-grained manner.
In particular, we will show how the data cube representation enables visual quality analysis using standard tools.

To put the reader into the context of this work, we introduce a use case:
\begin{quote}
Quimp is a startup company providing various life science linked datasets.
Their business model is that their customers (the real publishers) provide their data in various formats to the Quimp Portal.
Meanwhile Quimp semantically lift the data to a standardised RDF linked data representation using a number of pre-defined vocabularies.
This linked data is periodically updated, having new resources added and others becoming obsolete.
Quimp will then offer access for these linked datasets to data consumers.
Data consumers often complain about the fact that datasets suffer a lot from incorrect and inconsistent facts.
Concerned with these complaints, Quimp decided to start computing quality metrics on the semantically-enriched data.
This quality metadata is not only computed for one-time analyses, but it is stored, in order to track the long-term evolution of dataset quality across multiple versions.
Visualising the quality metrics and their development over time helps Quimp identify what aspects of their data are not up to standard, and therefore ensuring that quality over the different versions does not diminish.
\end{quote}


The remainder of this paper is structured as follows: in Section~\ref{sec:usecases} we discuss use cases for the daQ ontology, including its new multi-dimensional aspect. 
Then, in Section~\ref{sec:ontology} and \ref{sec:usage} we discuss the vocabulary design and show how data can be explored and used.
Finally, in Section~\ref{sec:relatedwork} we give an overview of similar ontology approaches before giving our final remarks in Section~\ref{sec:conclusion}.


\section{Use Cases}
\label{sec:usecases}
Linked Open Data quality has different stakeholders in a myriad of domains, however, the stakeholders can be cast under either \emph{publishers} or \emph{consumers}.
\emph{Publishers} are mainly interested in publishing data that others can reuse. 
\emph{Consumers}, both human end users and machine agents assisting them, require to use this published data in their applications.

Data consumers, both human and machine, may find it challenging to assess the quality of a dataset, i.e.\ its fitness for use. 
Currently, there is no standard way of how data publishers can assess the quality of their linked datasets.
Most of these publishers rely on their mutual trust they have with data providers, believing that the data they provide is good for data consumers, leaving the data publishers in the dark of the value and quality of their data.
Undoubtedly, quality metadata is also significant to the data consumer who ultimately has to decide what data is fit to their use case.
Activities of the recently formed \textit{W3C Data on the Web Best Practices Working Group} (DWBP)\footnote{\url{https://www.w3.org/2013/dwbp}}\footnote{The authors are affiliated with this WG, contributing to the standardisation of quality assessment of LOD.} include developing a standard vocabulary for assessing and representing metadata about quality.
Our daQ, or an adaptation or extension of it, is a candidate.

The following use cases (UC) show how both data publishers and consumers can benefit from having quality metadata about datasets.
The UC thus motivate the need for developing a standard like daQ.
Our original paper discusses further use cases~\cite{DebattistaEtAl:daQ:LDOW:2014}.

\subsection{UC1: Analysis of Data Versions}
Ideally, data publishers update their published datasets regularly to 
(a) keep the data fresh and up-to-date; 
(b) clean data to improve quality; 
(c) keep up with the data curation lifecycle.
However, it is sometimes difficult to identify which aspects of the data are lacking quality standards.
Furthermore, it is even more difficult to analyse how data quality changed over time.
Assuming that there are tools that automatically analyse data quality and output daQ metadata (such as our implementation mentioned in Section~\ref{sec:impl-eval}), their data cube structure enables a multi-dimensional representation, where the versions of a dataset form one dimension, and the different quality metrics form the other dimension.

\subsection{UC2: ``Fit'' Dataset for Retrieval}
\Ednote{CL@JD: I don't understand the section title.  Wouldn't “‘fit’ dataset\emph{s} for retrieval” or “retrieving ‘fit’ datasets” be easier to understand?}Alexander et al.~\cite{Alexander:LDOW09} provide the readers with a motivational use case with regard to how the VoID ontology (cf.\ Section~\ref{sec:relatedwork}) can help with effective data selection.
The authors describe how a consumer can find the appropriate dataset by: 
\begin{itemize}[nosep]
\item criteria related to content (what is the dataset mainly about);
\item its links from and to other datasets;
\item the vocabularies used in the dataset.
\end{itemize}
The daQ ontology gives an extra edge to ``appropriateness'' by providing the consumer with quality criteria on the candidate datasets.

An objective assessment of data quality enables data consumers to determine if a dataset is fit for a certain use case.
Currently, tools targeting human data consumers, such as semantic web search engines~\cite{Hogan2011:SWSE} or Data Web browsers~\cite{tbl:tabulator,Harth2010:Visnav,Heim2008:gfacet}, do not focus on dataset quality when presenting search results.
With the introduction of the daQ framework, tools that provide faceted browsing facilities, such as the CKAN data portal engine\footnote{\url{http://ckan.org}}, are enabled to provide more information about a dataset's quality attributes.
Such functionality is attributable to the flexibility of the ontology, enabling various filtering and ranking possibilities of the dataset quality metrics.  
This would permit human data consumers to better understand the quality attributes of a dataset, and thus choose which is the most fitting to their use case.
The recent addition of multi-dimensionality to quality metadata in the daQ model enables data consumers to track and follow quality improvements of data publishers on their datasets over time.
This also opens a sundry of opportunities leading to the assessment of data publishers regarding their willingness to enhance the value of the data in terms of quality.
To keep the quality metadata of open datasets easily accessible, we recommend that each dataset contains the relevant daQ metadata graph within the dataset itself.

\subsection{UC3: Link Identification}
Identifying links between existing datasets is one of the main drivers that makes the Linked Open Data cloud more coherent.
Tools such as LIMES~\cite{NGAU11} or Silk~\cite{vol09b} support the automatic identification of links according to built-in as well as user-defined criteria.
The introduction of quality metadata to datasets will add another criterion for link identification, in that linking algorithms can also take the quality of the target dataset into consideration before linking to it.
Linking tools could also consider the needs of a data consumer who might not only require to link to any high quality entity, but possibly even to those datasets which the consumer deems ``fit'' to her cause.
This can be done by ranking and filtering candidate datasets according to criteria such as weights on specific quality metrics defined by the consumer (as described in \textit{UC2}).
Linking resources of proven quality helps to improve the quality of both datasets participating in the link.
The generic framework proposed for the daQ ontology ensures that any custom metric defined by third parties can be easily integrated into any tool supporting such quality metadata for linking.

\subsection{UC4: Extension of the Five Star Scheme}
The popular five star scheme for deploying open data\footnote{\url{http://5stardata.info}}, which we propose to extend by a sixth star for quality, defines a set of widely accepted criteria that serve as a baseline for assessing data reusability.
The reusability criteria defined by the five star scheme and by the quality metrics are largely measurable in an objective way.
Thanks to such objective criteria, one can assess the reusability of any given dataset without the major effort of, for example, running a custom survey to determine whether its intended target audience finds it reusable.
Such a survey may, of course, still help to get an \emph{even better} understanding of quality issues.
As a consumer, the benefits of a sixth star is that good quality datasets can be discovered.
On the other hand, as a data publisher, the benefits of having the sixth star are that 
(i) the published data conforms to the established domain quality metrics; and 
(ii) catalogued and archived datasets (refer to~\cite{DebattistaEtAl:daQ:LDOW:2014}) can be easily discovered when consumers filter by quality aspects.


\section{The Dataset Quality Ontology (daQ)}
\label{sec:ontology} 
The idea behind the Dataset Quality Ontology\footnote{\url{http://purl.org/eis/vocab/daq}} (daQ) is to provide a generic core vocabulary, allowing a uniform definition of specific data quality metrics.
This metric definition would then allow publishers to represent the metadata resulting from benchmarking the quality of their datasets, or even to attach these metadata to their linked datasets.

\begin{figure*}[tbph]
\center
\includegraphics[page=1,trim=.2cm .8cm .1cm .795cm,clip,width=.85\textwidth]{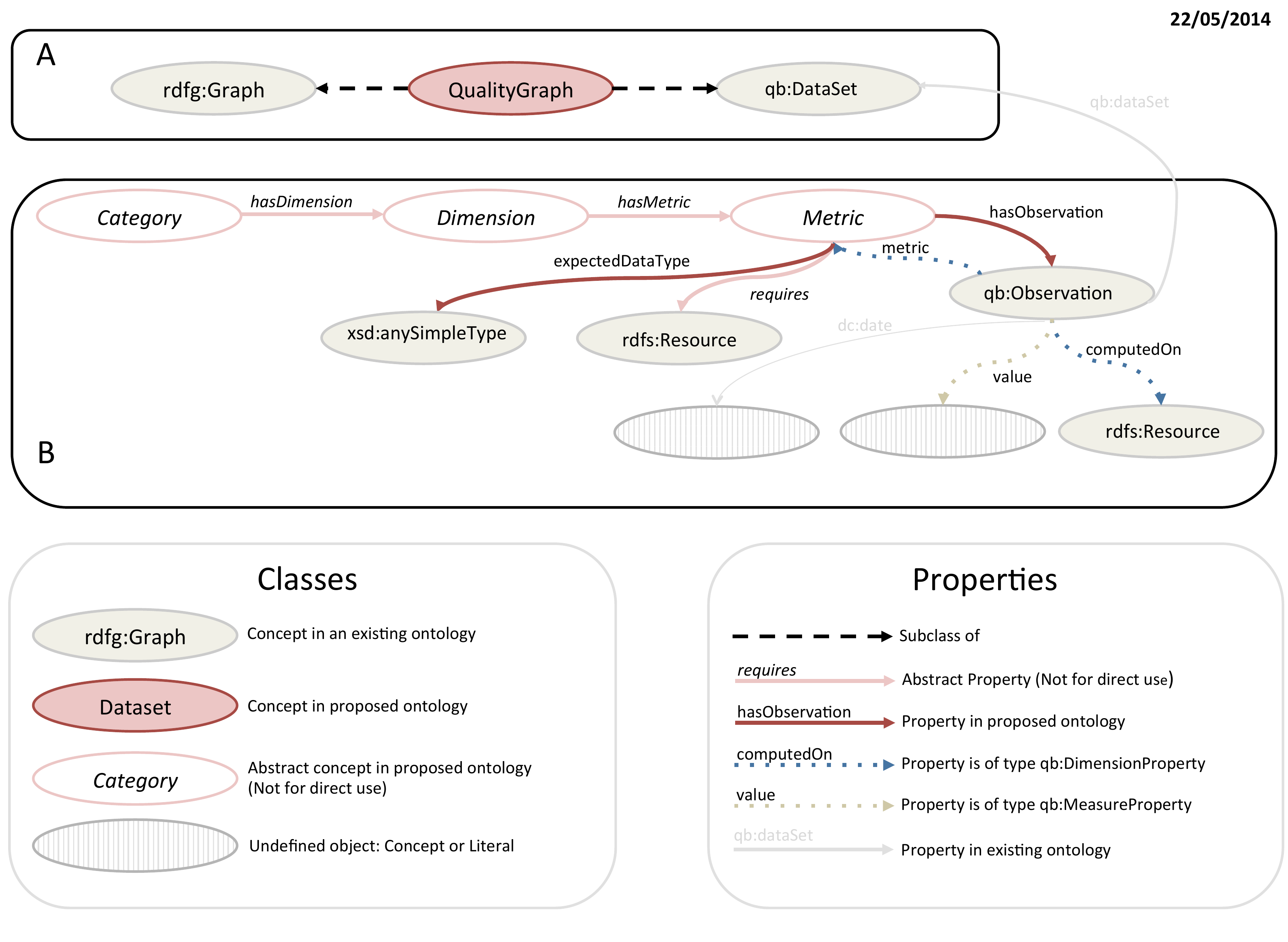} 
\caption{The extended Dataset Quality Ontology (daQ)}
\label{fig:daqExtended}
\end{figure*}		

\subsection{The basic daQ Concepts}
We first recapitulate the fundamental concepts, which were originally introduced in more detail in~\cite{DebattistaEtAl:daQ:LDOW:2014}. 
Quality metadata is intended to be represented as a \emph{Quality Graph}.
The latter concept is a subclass of \texttt{rdfg:Graph}~\cite{CBHS:NamedGraphs2005}.
This means that the quality metadata is stored and managed in a named graph that is separate from the dataset but can be embedded into the dataset if desired.
Named graphs also allow the digital signing of graphs~\cite{conf/semweb/CarrollBHS04}, thus ensuring trust in the computed metrics.

The daQ ontology distinguishes between three layers of abstraction, based on the survey work by Zaveri et al.~\cite{Zaveri2012:LODQ}.
As shown in Figure~\ref{fig:daqExtended} Box B, a quality graph comprises a number of different \emph{Categories}, which in turn possess a number of quality \emph{Dimensions}\footnote{In this paper we will refer to these as \emph{quality dimensions}, to avoid confusion with data cube dimensions}.
A quality dimension groups one or more quality \emph{Metrics}.

\subsection{Extending daQ for Multi-Dimension Representation and Statistical Evaluation}
The Data Cube Vocabulary~\cite{w3c:REC-vocab-data-cube-20140116}, abbreviated using the prefix \texttt{qb:}, allows the representation of statistical observations in multidimensional spaces.
In the initial daQ design~\cite{DebattistaEtAl:daQ:LDOW:2014}, computing quality metrics of different resources (usually: datasets), or even different revisions of the same resource, resulted in multiple \emph{quality graph}s, consisting of multiple instances of \textit{Metric} classes representing the individual observations.
Multidimensional analysis of these observations, e.g.\ across the revision history of a dataset, would thus have required complex querying.
Reusing the standardised Data Cube Vocabulary in daQ allows us to represent quality metadata of a dataset as a collection of \textit{Observation}.
\Ednote{CL@JD: It would make sense to have a data-cube-in-a-nutshell intro \emph{here}; that's why I'd put the following here (of course not mentioning additional dimensions, as we had not agreed on them).  IMHO for those readers who are not data cube experts it's \emph{not} redundant with what we say below.  ``The properties of an observation include \textit{Dimension Properties} (here: the quality metric computed, and the resource whose quality was assessed), and a \textit{Measure Property} (here: the metric value).''}
This also permits applying the wide range of tools that support data cubes to quality metadata, including the CubeViz visualisation tool\Ednote{CL@JD: In addition we should cite a publication. Can you find out the “right” one and tell me the name of the bib entry here?  Don't \textbackslash cite yourself as I have already hand-optimised the bib}\footnote{\url{http://cubeviz.aksw.org}}.

Figure~\ref{fig:daqExtended} shows the current state of daQ, where the introduction of data cubes entails some structural changes over the initial version of the ontology from~\cite{DebattistaEtAl:daQ:LDOW:2014}. 
A \emph{Quality Graph} is a special case of \texttt{qb:DataSet}, which allows us to represent a collection of quality observations complying to  a defined dimensional structure.
daQ defines the structure of such observations by the \texttt{qb:DataStructureDefinition} shown in Listing~\ref{lst:dsd_def}.

\begin{lstlisting}[float=tb,caption={The Data Structure Definition (Turtle Syntax)},label=lst:dsd_def, language=N3]
daq:dsd a qb:DataStructureDefinition ;
  # Dimensions: metrics and what they were computed on
  qb:component [
    qb:dimension daq:metric ;
    qb:order 1 ; ] ;
  qb:component [
    qb:dimension daq:computedOn ;
    qb:order 2 ; ] ;
  # Measures (here: metric values)
  qb:component [ qb:measure daq:value ; ] ;
  # Attribute (here: unit of measurement)
  qb:component [
    qb:attribute sdmx-attribute:unitMeasure ; 
    qb:componentRequired false ;
    qb:componentAttachment qb:DataSet ; ] .
\end{lstlisting}

The \texttt{daq:QualityGraph} definition in Listing~\ref{lst:qg_def} also specifies one restriction that controls the property \texttt{qb:structure} and its value to the mentioned definition, \Ednote{CL@JD: Please let's talk once more about how the following can be achieved: a data cube that has separate dimensions for “different datasets” and “different versions of datasets”, and metrics (with the same configuration) computed for all of them.  E.g.\ EFO-2013-Accessibility, EFO-2014-Accessibility, DBpedia-2013-Accessibility, DBpedia-2014-Accessibility. - JD@CL: Yes sure. I was also thinking about this lately - and I'm not entirely sure about it -, but first how about having some "real examples", as then we can see better how these are represented? The basic idea is explained though}thus ensuring that all \emph{Quality Graph} instances make use of the standard definition.
Having a standard definition ensures that all \emph{Quality Graph}s conform to a common data structure definition, thus datasets with attached quality metadata can be compared.

\begin{lstlisting}[caption={The Quality Graph Definition (Turtle Syntax)},label=lst:qg_def, language=N3]
daq:QualityGraph
  a rdfs:Class, owl:Class  ;
  rdfs:subClassOf rdfg:Graph , qb:DataSet , 
    [ a owl:Restriction ;
      owl:onProperty qb:structure ;
      owl:hasValue daq:dsd ];
  rdfs:comment "Defines a quality graph which will contain all metadata about quality metrics on the dataset." ;
  rdfs:label "Quality Graph Statistics" .
\end{lstlisting}

The detailed definitions of all abstract classes in Figure~\ref{fig:daqExtended} Box B, have been introduced in~\cite{DebattistaEtAl:daQ:LDOW:2014}.
The \texttt{daq:Metric} class now no longer has an immediate \emph{value} property, but is linked to from an \texttt{qb:Observation} by \texttt{daq:metric} (defined as a \texttt{qb:DimensionProperty}), whose inverse \texttt{daq:hasObservation} we define for convenience.
The properties \texttt{daq:computedOn} and \texttt{daq:value} are now defined as \texttt{qb:DimensionProperty} and \texttt{qb:MeasureProperty} respectively.
The former is defined \Ednote{CL@JD: Do you know that with qb:componentAttachment we could change this?  (This would avoid some redundancy, but on the other hand require more complex queries for certain analysis tasks.)}in each observation instance rather than once as a \emph{Quality Graph} property.
Each observation also has a \texttt{dc:date} property, which holds the timestamp of when it was computed.
Each custom metric definition should also include the \Ednote{CL@JD: really call it like this?  How about \textit{valueRange} instead?  (I think that's easier to understand, and it avoids confusion with the “expected datatype property” that the quality \emph{reports} for some metrics will have.)  Plus, should we also have properties for \textit{bestValue} and \textit{worstValue}? - JD: Lets talk about these later\\ CL: OK, we can move this to WWW.  So for now we'll go with this one property whose value is, IIUC, \emph{xsd:boolean}, \emph{xsd:double}, etc. – but I'm still a bit skeptical whether \emph{expected}DataType is a good name.}\texttt{daq:expectedDataType} property.
This will indicate the observation's value datatype.
The optional property \texttt{sdmx-attribute:unitMeasure} can be defined on an observation instance, enabling a system (application) to further specify the semantics of the unit of measurement of the value.
For example, some of a dataset's accessibility metrics are reasonably measured as durations in seconds, whereas the majority of metrics can be expressed as plain numeric figures in the normalised interval $[0,1]$.

\section{Using the Ontology}
\label{sec:usage} 

\subsection{Extending daQ}
\begin{figure*}[tbph]
\begin{center}
\includegraphics[width=\textwidth]{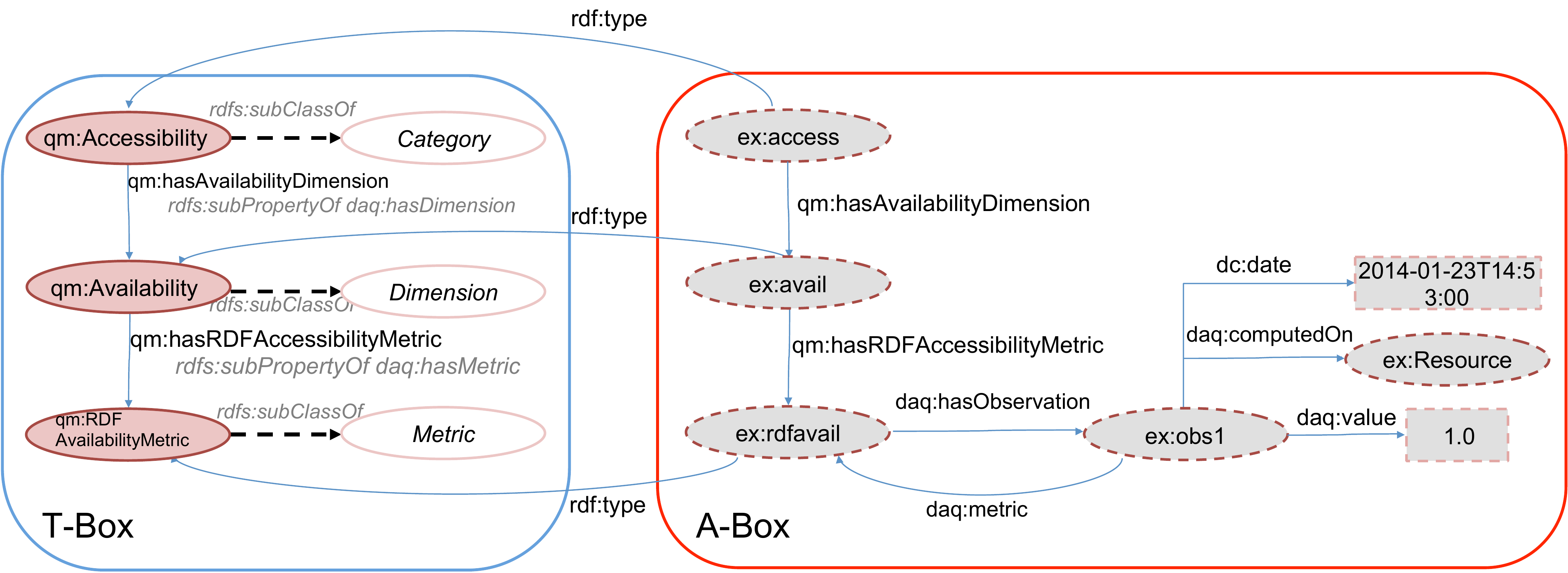}
\caption{Extending the daQ Ontology – TBox and ABox}
\label{fig:ext_daq}
\end{center}
\end{figure*}

The classes of the core daQ ontology are intended to be extended by specific and custom quality metrics that characterise a dataset's ``fitness for use''~\cite{Juran1974:biblatex} in a particular domain.
We have defined the quality dimensions and metrics described in~\cite{Zaveri2012:LODQ}, some of which are being considered to be standard metrics to calculate quality on Linked Open Data sets whilst others are specific to the DIACHRON project, in whose context we are doing this research (refer to Section~\ref{sec:diachron}).
\textbf{Extending} the daQ ontology means adding new quality protocols that inherit the abstract concepts (Category–Dimension–Metric).
Custom quality metrics do not need to be included in the daQ namespace itself; in fact, in accordance with LOD best practices, we recommend extenders to make them in their own namespaces.
Figure~\ref{fig:ext_daq} shows an illustrative example of extending the daQ ontology (TBox) with a more specific quality attribute – the RDF Availability Metric as defined in~\cite{Zaveri2012:LODQ} – and an illustrative instance (ABox) of how it would be represented in a dataset.

The \texttt{Accessibility} concept is defined as an \texttt{rdfs:subClassOf} the abstract \texttt{daq:Category}.
This category has five quality dimensions, one of which is the \textit{Availability} dimension.
This is defined as an \texttt{rdfs:subClassOf} \texttt{daq:Dimension}.
Similarly, \textit{RDFAvailabilityMetric} is defined as an \texttt{rdfs:subClassOf} \texttt{daq:Metric}\Ednote{CL@JD: OK, I give in :-)  Just one point: ``we would have Accessibility hasDimension Availability and Accessibility hasDimension ExtensionalConcisness'' – I will not argue that we should \emph{allow} this, but you need additional, more heavyweight semantics, to really \emph{prevent} people from saying ``(1) ex:myAccessibility dqm:hasAvailabilityDimension ex:myAvailability . (2) ex:myAccessibility dqm:hasAvailabilityDimension ex:myExtensionalConcisness .''  From ``dqm:hasAvailabilityDimension rdfs:range dqm:Availability'' and (2) we can first of all infer ``ex:myExtensionalConcisness rdf:type dqm:Availability''.  Suppose we know ``ex:myExtensionalConcisness rdf:type dqm:ExtensionalConciseness''.  So far it's still possible in RDFS.  We'd need the OWL axiom ``dqm:Availability owl:disjointWith dqm:ExtensionalConciseness'' to create an inconsistency here.  I.e. you'd have to add such pairwise disjointness axioms between all Metrics/Dimensions/Categories.  (And then, in a LOD setting, you can't be \emph{sure} that all users will have an OWL reasoner at hand; that's why many LOD vocabularies don't formally restrict all things that actually, I agree, don't make sense.)  But it's fine with me to go this way.  For encoding pairwise disjointness between more than 2 classes in RDF, check \url{http://www.w3.org/TR/owl2-mapping-to-rdf/} and search for ``DisjointClasses''.}.
The specific properties \textit{hasAvailabilityDimension} and \textit{hasRDFAccessibilityMetric} (sub-properties of \texttt{daq:hasDimension} and \texttt{daq:hasMetric} respectively) are also defined (Figure~\ref{fig:ext_daq})

\subsection{Publishing daQ Metadata Records}
\label{sec:publ-daq-metad}
We encourage publishers of linked \emph{open} data to offer daQ metadata as an RDF named graph in their published dataset.
Since such a daQ metadata record requires a lot of metrics to be computed, it is not normally intended to be authored manually.
Publishing platforms such as CKAN should offer such an on-demand computation to dataset publishers (as sketched in~\cite{DebattistaEtAl:daQ:LDOW:2014}; see section~\ref{sec:impl-eval} for our own implementation progress).
Listing~\ref{lst:listing1} shows an instance of the \texttt{daq:QualityGraph} in a dataset.
\textit{ex:qualityGraph1} is a named \texttt{daq:QualityGraph}.
The defined graph is automatically a \texttt{qb:DataSet}, and \Ednote{CL@JD: If we want to emphasize this restriction, why do we redundantly mention the DSD in the listing?  Oh BTW getting back to our email conversation where I complained about this DSD restricting the user.  Would it be possible to say that quality metadata must \emph{at least} have the dimensions computedOn and metric, but \emph{may} have further dimensions? - JD: i don't think that is possible without "overriding" the defined datastucture – CL: OK, this is another pro/con thing we may want to touch on in the ``formalisation'' section of WWW}due to the restriction placed on the \texttt{daq:QualityGraph} (see Listing~\ref{lst:qg_def}), the value for the \texttt{qb:structure} property is defined as \texttt{daq:dsd} (see Listing~\ref{lst:dsd_def}).
In the named graph, instances for the \texttt{daq:Accessibility}, \texttt{daq:Availability}, \texttt{daq:EndPointAvailabilityMetric} and \texttt{daq:RDFAvailabilityMetric} are shown.
A metric instance may have been used to make a number of observations.
Each of these observations specifies the metric value (\texttt{daq:value}), the resource the metric was computed on (\texttt{daq:computedOn}), when it was computed (\texttt{dc:date}), the metric instance (\texttt{daq:metric}) and finally to what data cube the observation is defined in (\texttt{qb:dataSet}).

\subsection{Exploring and Visualising the daQ Metadata}
\begin{figure}[ht]
\centering
\subfigure[Horizontal Bar Chart]{
  \includegraphics[scale=.25]{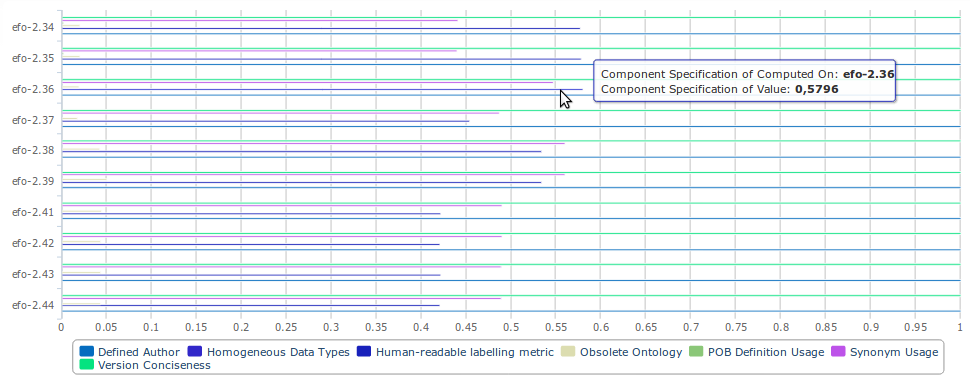}
  \label{fig:hor_chart}}
\quad
\subfigure[Vertical Bar Chart]{
  \includegraphics[scale=.25]{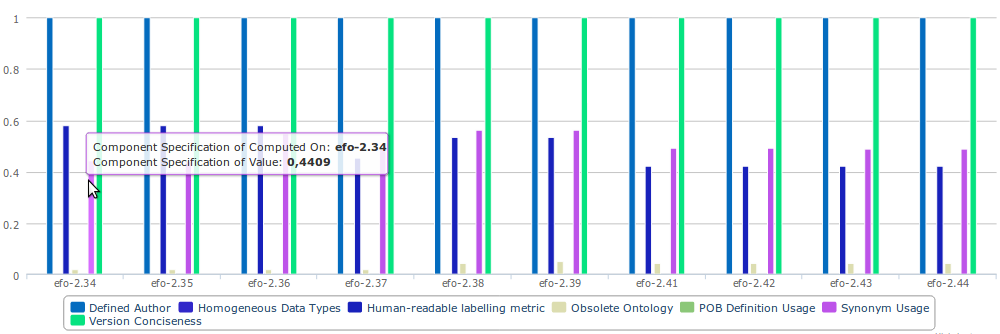}
  \label{fig:ver_chart}}
\subfigure[Radar Chart]{
  \includegraphics[scale=.25]{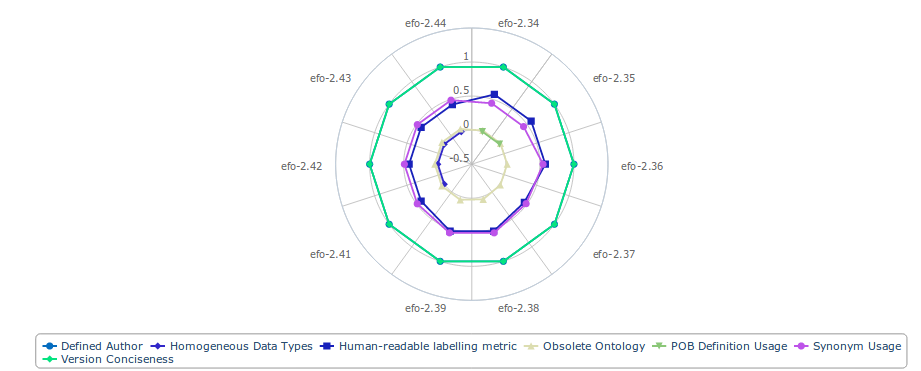}
  \label{fig:rad_chart}}
\quad
\subfigure[Lines Plot]{
  \includegraphics[scale=.25]{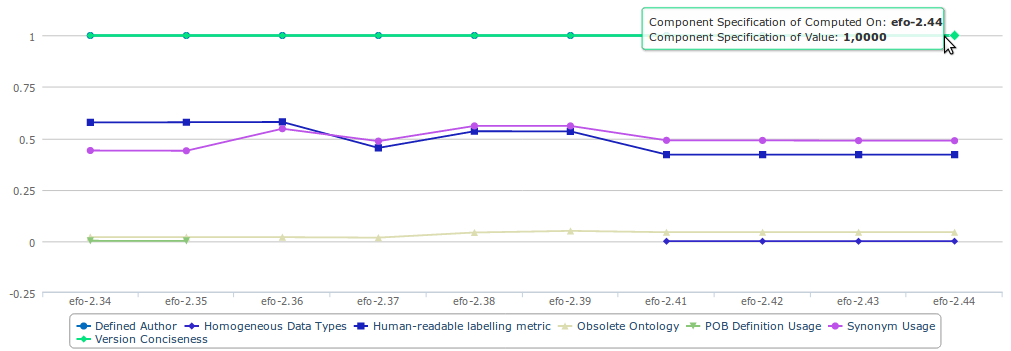}
  \label{fig:line_chart}}
\caption{Visualising Quality Metadata}
\label{fig:visuals}
\end{figure}

CubeViz is a tool for visualising data cubes.
\Ednote{CL@JD: If we end up with a lot of free space, we can make this figure span two columns and put it on a page of its own.}Figure~\ref{fig:visuals} depicts four different CubeViz chart visualisations of the same quality metadata\footnote{The quality metadata used can be found in \url{https://raw.githubusercontent.com/diachron/quality/master/src/test/resources/cube_qg.trig}}.

A \emph{horizontal bar} represents each metric (Figure~\ref{fig:hor_chart}) and shows its value (x-axis) with respect to the \Ednote{CL@JD: Old comment (maybe even wrong; I think I'm now understanding this ``version'' dimension differently); I rephrased “version of the document” into the more neutral “dataset”, as “version” somehow suggests that we have “time” as a dimension in the data cube, which we don't, at least not here. - JD: we have 2 dimensions: dataset (irrelevant if it is the same dataset with different versions or different datasets in the same domain) and metric... the paper is more about the vocabulary.. its usage was secondary.. if we talk about its usage then we have to come up with more use cases – CL: OK, maybe my concern was from the time when I thought that this paper would \emph{only} be about data cube aspects.  We might leave this to WWW, as another point of ``discussing our design decisions'' (rather than merely presenting them)}dataset (y-axis).
Here, the different “datasets” analysed are actually successive revisions of one dataset.
This chart provides a clear view of how the value associated to each one of the measured metrics changes as the dataset (in this case) evolves.
The horizontal layout \Ednote{CL@JD: for now we can make such claims; in fact no reviewer complained about them.  However in future these would be things to prove by a usability evaluation.}is appropriate when the range of metric values is wide, and the number of different datasets is relatively small.

Similar to the horizontal bars chart, the \emph{vertical bar chart} (Figure~\ref{fig:ver_chart}) allows the user to compare the values computed for each of the metrics (y-axis), with respect to the dataset (x-axis).
In contrast with its horizontal counterpart, this chart is more appropriate when there are many datasets analysed but the range of metric values is not so wide.

In the \emph{radar chart} (Figure~\ref{fig:rad_chart}), the datasets are represented as slices of a circle and the values corresponding to the metrics are depicted as points and lines of a particular colour.
This chart provides a clear view of how the values of the metric differ from each other for each particular dataset. 
Furthermore, it allows one to assess the overall quality of a dataset, by showing whether the values of the metrics are concentrated around sections of the circle regarded as “good” or “bad”.

The lines plot (Figure~\ref{fig:line_chart}), lists the different datasets against the values of the metrics.
Here, where “different datasets” are actually different revisions in the evolution of one dataset, this plot provides a comparison of the evolution of the quality of the dataset, with respect to each metric.
The lines emphasise the points where the values of the metrics changed noticeably from one version to the next.

\subsection{Analysing Observations based on Quality Dimensions}
The daQ framework allows the definition of quality metrics in three levels of abstraction: Category–Dimension–Metric.
Although the instances have a link between these three levels, we only perform observations on the metric level.
Therefore, when visualising and analysing observations, the consumer would only be able to observe the metrics from all quality categories and dimensions, instead by specific quality dimension or category.
Thanks to the link between the three levels of abstraction, no manual human intervention is required to analyse a set of metrics \emph{grouped} by a specific quality dimension.

Data Cube slices allow the grouping of observation subsets.
Since slices are not intended to represent arbitrary selections in a data cube, but only selections that result from fixing the values for some dimensions, \texttt{qb:ObservationGroup} has to be used instead.
\Ednote{CL@JD: given that the CONSTRUCT only uses ?obs, the SELECT doesn't have to output ?metricInst, right?}Listing~\ref{lst:listing5} shows a SPARQL CONSTRUCT defining an observation group, where all observations in the Accessibility dimension are grouped in a constructed \texttt{ex:dimObs1} resource.\Ednote{CL@JD: Once more I think this is a good example for not having hasDimension/hasMetric as abstract properties.  If they were concrete properties, this query would have been easier to write. JD@CL: true but then any metric is any dimension and any category which is wrong}\Ednote{CL@JD: Maybe obsolete, will check later: I think selecting observations and filtering them by their daq:metric dimension is more intuitive to understand for people who know data cubes. JD@CL: this is meant to be technical - we cannot say filter by dimension without first doing the observation groups - therefore i needed to explain that we cannot do slices but observation groups (which in turn would enable the filtering by dimension/category etc...)}
\begin{lstlisting}[caption={Creating A Data Cube Observation Group using SPARQL},label=lst:listing5, language=SPARQL,escapechar=@]
CONSTRUCT {
  ex:dimObs1 a qb:ObservationGroup ;
    qb:observation ?obs . 
}
WHERE {
 	SELECT DISTINCT ?metricInst ?obs {
	?dimInst	a	dqm:Accessibility .
	?dimInst	?prop	?metricInst .
	?metricInst daq:hasObservation	?obs .
	?metricInst	a	?metric .
	
	GRAPH <http://www.diachron-fp7.eu/dqm@\#@> {
		?prop	rdfs:subPropertyOf	daq:hasMetric .
		?metric	rdfs:subClassOf	daq:Metric .
	}
}
\end{lstlisting}
The resulting construct output is shown in Listing~\ref{lst:listing6}.
\begin{lstlisting}[caption={A Data Cube Observation Group.},label=lst:listing6, language=N3]
ex:dimObs1 a qb:ObservationGroup ;
 qb:observations ex:obs1 , ex:obs2 , ex:obs3 , ex:obs4 .
\end{lstlisting}

\begin{lstlisting}[basicstyle=\footnotesize \ttfamily,caption={A \emph{Dataset Quality Graph}},label=lst:listing1, language=N3]
# ... prefixes
# ... dataset triples

ex:qualityGraph1 a daq:QualityGraph ;
	qb:structure daq:dsd . 
	
ex:qualityGraph1 {

	# ... quality triples
	ex:accessibilityCategory
		a dqm:Accessibility ;
		dqm:hasAvailabilityDimension
			ex:availabilityDimension .
	
	ex:availabilityDimension
		a dqm:Availability ;
		dqm:hasEndPointAvailabilityMetric
			ex:endPointMetric ;
		dqm:hasRDFAvailabilityMetric
			ex:rdfAvailMetric .
	
	ex:endPointMetric
		a dqm:EndPointAvailabilityMetric ;
		daq:hasObservation ex:obs1, ex:obs2 .
			
	ex:obs1 a qb:Observation ;
		daq:computedOn <efo-2.43> ;
		dc:date "2014-01-23T14:53:00"^^xsd:dateTime ;
		daq:value "1.0"^^xsd:double ;
		daq:metric ex:endPointMetric ;
		qb:dataSet ex:qualityGraph1 .
	
	ex:obs2 a qb:Observation ;
		daq:computedOn <efo-2.44> ;
		dc:date "2014-01-25T14:53:00"^^xsd:dateTime ;
		daq:value "1.0"^^xsd:double ;
		daq:metric ex:endPointMetric ;
		qb:dataSet ex:qualityGraph1 .
	
	ex:rdfAvailMetric
		a dqm:RDFAvailabilityMetric ;
		daq:hasObservation ex:obs3, ex:obs4 .
				
	ex:obs3 a qb:Observation ;
		daq:computedOn <efo-2.43> ;
		dc:date "2014-01-23T14:53:01"^^xsd:dateTime ;
		daq:value "1.0"^^xsd:double ;
		daq:metric ex:rdfAvailMetric ;
		qb:dataSet ex:qualityGraph1 .
	
	ex:obs4 a qb:Observation ;
		daq:computedOn <efo-2.44> ;
		dc:date "2014-01-25T14:53:01"^^xsd:dateTime ;
		daq:value "0.0"^^xsd:double ;
		daq:metric ex:rdfAvailMetric ;
		qb:dataSet ex:qualityGraph1 .
		
	# ... more quality triples
}
\end{lstlisting}



\section{Implementation, Evaluation, and the DIACHRON Project}
\label{sec:impl-eval}
\label{sec:diachron}

We are currently implementing the following support for automated quality assessment and analytics: (i) Luzzu, a generic quality assessment framework based on the daQ ontology, implemented in Java using the Jena RDF libraries and offering a web service interface\footnote{\url{https://github.com/EIS-Bonn/Luzzu}}, and (ii) so far 38 concrete quality metrics over linked open datasets written by ourselves or by collaborators in the DIACHRON project.
They are described in an extension of the daQ ontology\footnote{data quality metrics (DQM; \url{http://purl.org/eis/vocab/dqm})} and implemented on top of the Luzzu framework\footnote{\url{https://github.com/diachron/quality}}.

The DIACHRON project (``Managing the Evolution and Preservation of the Data Web''\footnote{\url{http://diachron-fp7.eu}}) combines several of the use cases mentioned so far in the application domains of open data, enterprise intranets and scientific data.
DIACHRON's central cataloguing and archiving hub is intended to host datasets throughout several stages of their life-cycle (cf.~\cite{Auer+ISWC-2012:biblatex} for a general introduction to the linked data life-cycle), mainly evolution, archiving, provenance, annotation, citation and quality assessment.
Beyond the implementation currently in progress, we will, together with the project partners, implement a web frontend that pulls dataset metadata from installations of the CKAN data portal system in order to
\begin{itemize}
\item allow data publishers to perform quality assessment on datasets, including the generation of quality metadata, visual analytics and support with cleaning datasets that are affected by quality problems, and to
\item allow data consumers to filter and rank datasets by multiple quality dimensions.
\end{itemize}
The daQ ontology is the common language understood by these components.

Further progress with this implementation will enable us to evaluate daQ and its concrete extensions with regard to questions such as how much time it takes to compute quality metrics for big datasets, to what extent the ontological foundation facilitates the agile development of end-user frontends, how well our visualisation of multi-dimensional quality metadata enables publishers to improve the quality of their data, and how well quality-based dataset filtering and ranking supports consumers in finding high-quality data.


\section{Related Work}
\label{sec:relatedwork} 

To the best of our knowledge, the Data Quality Management (DQM) vocabulary~\cite{Furber2011:TVD} is the only one comparable to our approach. 
F\"{u}rber et al.\ propose an OWL ontology that primarily represents data requirements, i.e.\ what quality requirements or rules should be defined for the data.
Such rules can be defined by the user herself, and the authors present SPARQL queries that ``execute'' the definitions of the requirements  to compute metrics values.
Unlike our daQ model, the DQM defines a number of classes that can be used to represent a data quality rule.
Similarly, properties for defining rules and other generic properties such as the rule creator are specified. 
The daQ model allows for integrating such DQM rule definitions using the \emph{daq:requires} abstract property, but we consider the definition of rules out of daQ's own scope.
Also, the proposed daQ ontology gives the freedom to the user to define and implement any metrics required for a certain application domain.

\Ednote{CL@JD: I know you are proud of this, but as it is, to my understanding, not yet a widely established ontology engineering methodology, but just an approach that you followed both in that paper and here, I think we can survive without mentioning it (should space become scarce).}Our design approach is inspired by the digital.me Context Ontology (DCON\footnote{\url{http://www.semanticdesktop.org/ontologies/dcon/}})~\cite{Attard2013:OSR}.
Attard et al.\ present a structured three-level representation of context elements (Aspect-Element-Attributes).
The DCON ontology instances are stored as Named Graphs in a user's Personal Information Model.
The three levels are abstract concepts, which can be extended to represent different context aspects in a concrete ubiquitous computing situation.

Ermilov et. al~\cite{ermilov-2013-kesw} present a framework calculating comprehensive statistics on Linked Open datasets.
This statistical metadata has an underlying ontology based on VoID (see below) and Data Cube.
The authors argue that since the VoID ontology was not sufficient to cover the required statistical concepts, the Data Cube extension was required.
This extension also presented an opportunity to the authors to represent such statistical data using arbitrary attribute dimensions.
Motivated by this work, we model resources and their calculated metrics as Data Cube observations, allowing us to represent quality metadata in a multi-dimensional manner.\Ednote{CL@JD: found this by googling “data quality” and “data cube”: \url{http://ieeexplore.ieee.org/xpl/login.jsp?tp=&arnumber=6199204&url=http://ieeexplore.ieee.org/xpls/abs_all.jsp?arnumber=6199204} looks promising (Multidimensional traffic GPS data quality analysis using data cube model).  No \emph{RDF} data cubes, but at least data cubes for analysing data quality.  Can't access it right now (but on Monday latest); would be good if you could have a look into it.}
 
The W3C recommends VoID and the Data Catalog Vocabulary (DCAT~\cite{w3c:REC-vocab-dcat-20140116}) for metadata describing datasets.
The ``Vocabulary of Interlinked Datasets'' (VoID) ontology allows the high-level description of a dataset and its links~\cite{Alexander:LDOW09,w3c:NOTE-void-20110303}.
On the other hand, DCAT describes datasets in data catalogs, which increase discovery, allow easy interoperability between data catalogs and enable digital preservation.
With the daQ ontology, we aim to \Ednote{CL@JD: Reviewer 3 wants us to say more explicitly why we didn't reuse/extend VoID.  My superficial intuition: nothing in VoID was really applicable to our specific needs of talking about quality metrics.  But would be good to have one well-written sentence on this.}extend what these two ontologies have managed to achieve for datasets in general to the specific aspect of data \emph{quality}: enabling the discovery of a good quality (fit to use) datasets by providing the facility to ``stamp'' a dataset with quality metadata.



\section{Concluding Remarks}
\label{sec:conclusion}
We presented the Dataset Quality Ontology (daQ), a core vocabulary for representing quality benchmarking metadata of linked open datasets, which makes use of the Data Cube Vocabulary.
In Section~\ref{sec:usecases} we presented a number of use cases that motivated our idea.
These included analysis of data versions, dataset retrieval, automatic link identification based on the quality of data entities, and finally the extension of the five star open data scheme by a star for quality.
The precise definition of these use cases assisted in the development of the daQ ontology, which involved reuse of the Data Cube (Section~\ref{sec:ontology}).

The ontology is progressing in a fast pace, and further developments to cover the intended use cases are also in the pipeline.
The next iteration phase is to further model the daQ ontology to cover the provenance aspect of quality metadata.
The development and extension of new concepts to the daQ ontology should ensure that (i) high standards are kept, and (ii) that the ontology is not bloated out of proportion – i.e. keeping with the main idea of a light-weight quality assessment framework.

Currently, using daQ, we are in the process of implementing (Sections~\ref{sec:usage} and \ref{sec:diachron}) a number of domain-specific and domain-independent metrics, following a survey of linked data quality metrics~\cite{Zaveri2012:LODQ}.
As quality metadata describes the dataset on which quality was calculated, it makes sense to maintain it close to the dataset.
In LOD settings we consider it most reasonable to embed the quality metadata into the dataset itself, which is possible thanks to the named graphs approach we chose.

We also demonstrated specific advantages of basing daQ on the Data Cube Vocabulary: quality metadata can be visualised using available Data Cube enabled applications such as CubeViz, and observations can be grouped together automatically using the daQ three level abstract layer.

One of the tools which will support the daQ framework is the DIACHRON platform.
This platform will enable consumers to rank and filter datasets by quality.
Having tools and platforms supporting the daQ will finally allow us to test and evaluate the ontology thoroughly, to see whether the daQ (and the quality metadata) itself is of a high quality, i.e.\ fit for use.


\section{Acknowledgments}
This work is supported by the European Commission under the Seventh Framework Program FP7 grant 601043 (\url{http://diachron-fp7.eu}).
\Ednote{CL@JD: I think something like this is appropriate; please check}We would like to thank Santiago Londoño, who carried out the visual analytics of our sample quality metadata.

%
\printbibliography
%
%
\end{document}
